\documentclass[aps,prl,twocolumn]{revtex4}
\usepackage{graphicx}
\begin{document}

\maketitle

\noindent {\bf Comment on ``Time-Dependent Density-Matrix
Renormalization Group: A Systematic Method for the Study of
Quantum Many-Body Out-of-Equilibrium Systems'' }

In a recent Letter\cite{cm}, Cazalilla and Marston (CM) proposed a
time-dependent density-matrix renormalization group (TdDMRG)
algorithm for accurate evaluation of out-of-equilibrium properties
of quantum many-body systems. An interesting result they obtained
is that for a point contact junction between two Luttinger liquids,
a current oscillation develops after initial transient in the
insulating regime (Fig. 2 in Ref. 1). They attributed this
oscillation to a non-linear response of the Luttinger liquid. Here
we would like to point out that (a) the oscillation they observed
is an artifact of the method; (b) the TdDMRG can be significantly
improved by extending the definition of the density matrix to
adapt the non-equilibrium evolution of the ground state.

In the TdDMRG scheme of CM, it is the equilibrium ground state
that is targeted and used in the construction of the reduced
density matrix. At the beginning of the evolution, the state does
not deviate much from the ground state $|\Psi_0\rangle$, and
results obtained with that approach are accurate. However, the time
evolution of the ground state evokes excited states in a
non-equilibrium system, so the long time behavior of the state
becomes very poor and could not be substantially improved by
keeping more states. This can be clearly seen from Fig. 1, where
the numerical results obtained with that approach (i.e. the
$N_t=0$ curves) for the two models in
Ref. \onlinecite{cm} begin to deviate from the exact ones at
$t\sim 25$.

\vspace{-0.7cm}
\begin{figure}[h]
\includegraphics[width = 7cm, angle = 0]{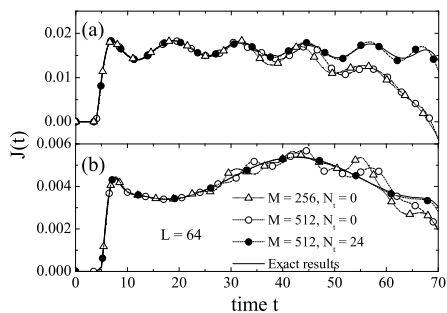}
\vspace{-0.7cm}
\caption{Current for (a) a quantum dot and (b) a junction with $V=0$,
defined by Eqs. (2) and (5) in Ref. \onlinecite{cm}, respectively.
The parameters are the same as in Ref. \onlinecite{cm}.} \label{fig1}
\end{figure}
\vspace{-0.2cm}
To retain the information on the relevant excitation states, we
propose to define the density matrix from the time dependent
wavefunction $|\Psi(t)\rangle$ as
\begin{equation}
\rho = \sum_{i =0}^{N_t} \alpha_i |\Psi(t_i)\rangle\langle
\Psi(t_i)|, \qquad \sum_i \alpha_i = 1,
\end{equation}
where$|\Psi(t_0)\rangle=|\Psi_0\rangle$, $t_0$ is the starting
time at which a bias is imposed and $N_t$ is the number of equal
time intervals within the whole evolution time to be evaluated.
The reduced density matrix for the system is defined by tracing
over the degrees of freedom of the environment. Both the time
evolution of the wavefunction and the newly defined reduced
density matrix should be evaluated from the
beginning of the DMRG truncation, not just for the final lattice
system. The density matrix used by CM corresponds to the $N_t=0$
case. In our calculation, we took $\alpha_i = 1/2$ for $i=0$ and
$1/2N_t$ otherwise. Empirically, we find that $N_t\ge 6$ is
generally needed. Fig. 1 shows our results with $N_t=24$.
By comparison with the exact results, we find that this new scheme
significantly improves the large time scale behavior of the current.
\vspace{-0.7cm}
\begin{figure}[h]
\includegraphics[width = 7cm, angle = 0]{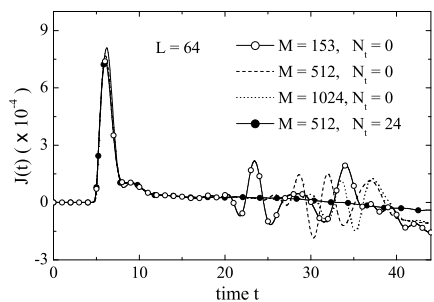}
\vspace{-0.7cm} \caption{Tunnelling current for the model defined
by Eq. (5) in Ref. [1] with $V = 1.1 w$ and $\delta
\mu=6.25\times10^{-2}w$.} \label{fig2}
\end{figure}

\vspace{-0.2cm} Now let us turn to the tunnelling junction between
two Luttinger liquids in the insulating regime. For $N_t=0$, our
calculation confirms the existence of a current oscillation around
$t\sim 20$. However, we find that the oscillation depends on the
number of states retained. By further applying the new TdDMRG
scheme to this system, we find that the oscillation does not exist
at all. In Fig. 2, the numerical results with $N_t=0$ and $N_t=24$
are shown for detailed comparisons. Therefore, the current
oscillation observed by CM is in fact an artifact of the
TdDMRG\cite{cm}, rather than a nonlinear response of the system to
the applied bias. This also shows that in the study of physical
properties of a quantum state out of equilibrium, it is important
to include the relevant excitation states in the definition of the
reduced density matrix.

\vfil
\vspace{0.2cm}
\vspace{0.1cm}

 \noindent H.G. Luo$^{1,2}$,T. Xiang$^{1}$, and X.Q.  Wang$^{1}$

\vspace{0.15cm}
\noindent $^1$Institute of Theoretical Physics\\
Chinese Academy of Sciences\\
P.O.Box 2735, Beijing 100080, China

\vspace{0.1cm} \noindent $^2$Department of Modern Physics\\
 Lanzhou University\\
 Lanzhou 730000, China

\end{document}